# Rapid-scanned and self-corrected repetition rates enabled in a bidirectional polarization-multiplexed fiber laser

Bowen Liu,[1] Maolin Dai,[1] Takuma Shirahata,[1,2] Yifan Ma,[1] Shinji Yamashita,[1,2] and Sze Yun Set[1,2,*]

[1]Department of Electrical Engineering and Information Systems, the University of Tokyo, 7-3-1 Hongo, Bunkyo-ku, Tokyo 113-8656, Japan
[2]Research Center for Advanced Science and Technology, the University of Tokyo, 4-6-1 Komaba, Meguro-ku, Tokyo 153-8904, Japan
*set@cntp.t.u-tokyo.ac.jp

**Abstract:** Repetition-rate-scanned lasers are practical in "accordion" frequency comb generation that serves as a variable gearbox connecting optical and radio wave domains. Rapid and wide-range scanned repetition rate can benefit versatile purposes, however scanning robustness remains unsecured that typically requires complicated feedback loops. Recently, multiplexed lasers have been demonstrated with the nature of common-noise rejection among simultaneously emitted combs. Here, we propose a bidirectional polarization-multiplexed fiber laser that delivers synchronized pulses with rapid-scanned and reference-free repetition rates. Benefiting from the all polarization-maintaining fiber configuration, the laser shows good robustness and inter-comb coherence. As rapid as 493.5 kHz/s scanning rate over 329-kHz scanning range of fundamental repetition rate is realized. The 1-hour and 1-day maximal variations of difference frequency are merely 0.52 Hz and 5.46 Hz. The capability to rebuilt steady state after mode hopping is also demonstrated. These results provide a promising solution for developing high-performance accordion-frequency laser sources.

1. Introduction

Pulsed lasers emit pulses at a constant repetition rate that corresponds to the reciprocal of the temporal pulse spacing. Particularly, a mode-locked fiber laser delivers tens of megahertz to even a few gigahertz repetitive pulses that plays an important role in filling the technical gap between optical and radio waves [1,2]. The evenly spaced pulses generated in an ideally stabilized mode-locked laser are mapped into an optical frequency comb with terahertz-frequency teeth that are separated by a pulse-repetition frequency. Computing and communication technologies relying on radio frequencies thus are seamlessly linked to $10^4$-times higher optical frequencies, greatly benefiting applications in timekeeping, precise ranging and green gases tracking [3-6]. Furthermore, a repetition-rate-scanned laser is practical for the generation of "accordion" optical frequency comb, which serves as a gearbox to down converse high optical frequencies into low radio frequencies in a tunable manner [2,7]. The attractive potential of a rapidly repetition-rate-scanned laser is clear, such as a versatile reference laser to fast match different combs [8], or a heterodyne meter requires no expensive optical spectral analyzers [9]. In 2004, B. R. Washburn *et al.* reported a phase-locked, self-referenced frequency comb with scanned repetition rates, and the fastest scanning speed achieved was 19.91 kHz/s [7]. Similar results were also demonstrated in a figure-9 laser scanning over ~800 kHz in 2010 [10]. Switchable high repetition rates up to 4 GHz were achieved based on harmonic-injection frequency multiplication incorporating a master laser and a slave laser [11]. Recently, it has also been reported that the repetition rate of a Mamyshev oscillator can vary within hundreds of megahertz depending on the pump power. [12]. Besides, harmonic mode-locking can also be used to quickly switch repetition rates over a wide range [13,14]. Although these results demonstrate the feasibility of rapidly changing repetition rates over a large dynamic range, a tunable delay line is still a favorable solution due to continuity concerns.

However, the repeatability and robustness has become general problems for repetition-rate-scanned lasers. Servos for real-time frequency correction are necessary, and self-referencing systems are critical for the precise positioning of the frequency teeth [15]. Alternately, dual-comb lasers have been developed to generate reference-free difference frequencies and greatly increase the sampling speed [16]. Recently, the generation of free-running dual combs in a single oscillator has attracted intensive attention. Cancellation of common-mode noise derived from the shared cavity secures highly stable difference frequencies, which could reduce the complexity and cost of dual-comb systems. Employing different waveguide properties of optical fibers, single-oscillator lasers support multiplexing of pulses such as wavelength-multiplexing, thereby emitting asynchronous dual-repetitive pulse trains. In 2016, X. Zhao *et al.* proposed the free-running dual-comb generation in a wavelength-multiplexing mode-locked fiber laser with only 16-mHz standard deviation (SD) regarding the frequency fluctuations [17]. The ability of wavelength-multiplexing lasers to suppress common-mode phase noise has also been experimentally confirmed [18]. This year, we reported a hybrid-tunable laser that provides stable dual-wavelength emission and flexible switching of single/dual-comb operation [19]. Whereas, generating asynchronous dual-repetitive pulse trains in a wavelength-multiplexing laser has less good coherence and limited bandwidth. To this end, configurations of bidirectional-propagating [20-24] and polarization-multiplexing [25,26] are also borrowed from optical communication into laser design. T. Ideguchi group reported a bidirectional dual-comb solid laser that experiences 0.1-Hz fluctuation SD with adjustable repetition rates [20]. N. Nishizawa *et al.* demonstrated a bidirectional polarization-maintaining fiber laser with 1-Hz maximum change in difference repetition rate during 7-hour free-running [21]. K. Minoshima team has developed both bidirectional and polarization-multiplexed fiber lasers, which are characterized with low Allan deviations and servo-free heterodyne detection [22,25]. In particular, a polarization-dependent bidirectional isolator called NINJA was proposed lately, which further improves polarization isolation and enhances controllability of dual combs in a bidirectional polarization-multiplexing fiber laser [26].

In this paper, a bidirectional polarization-multiplexes mode-locked fiber laser is proposed for rapidly scanned repetitive pulsing with reference-free repetition rates. The laser consists of a main oscillator and two auxiliary branches, where two orthogonally polarized pulse trains — a scan and a local comb, are simultaneously emitted and travel in opposite directions. Around 500-kHz/s scanning rate is realized by a delay line driven by resonant piezoelectric motors. Attributing to the common time jitters in the bidirectionally shared oscillator, the repetition rates of the two combs can remain synchronized with the ability to rebuild steady state. Thus, the robustness of difference frequency is secured and the repetition rate of either comb is reference-free. This work demonstrates an engineering blueprint for developing high-performance repetition-rate-scanned light sources. As it is mainly based on commercially available solutions, the results are repeatable and optimizable according to general uses.

2. Laser setup

The proposed fiber laser is illustrated as shown in Fig. 1, which is composed of a shared bidirectional cavity and two independent one-way branches. The 980-nm pump light from the laser diode (LD) is first split by a 50:50 optical coupler (OC3), and then coupled from both sides of a 1.4-m PM erbium-doped fiber (PM-EDF) through two fused wavelength-division multiplexers (WDM1, WDM2). The two outputs of OC3 are spliced to the pump ports of WDM1 and WDM2 at 0 and 90 degrees, respectively. Thus, the PM-EDF is bidirectionally pumped to provide the shared cavity with polarization-orthogonal gain. To connect the main oscillator with the two branches, a polarizing beam splitter (PBS) and a polarization-independent circulator (PI-CIR) are employed to guide the emitted light to propagate either clockwise or counterclockwise depending on its polarization. Here, the clockwise light (blue marked) travels along the slow axis of panda fibers: injecting from port-2 and output from port-3 of the PI-CIR, passing through a tunable attenuator (Att.) and a saturable absorber (SA1), then injecting from port-s and coupling back to the shared cavity from port-c of the PBS. Therefore, the total length of fibers in clockwise direction is ~7.00 m, corresponding to a net-dispersion of around -0.150 ps$^2$. On the other hand, counter-clockwise light (red marked) propagates along the fast axis: injecting from port-c and output from port-f of the PBS, passing through the SA2 and a tunable delay line, finally injecting from port-1 and returning to the main oscillator from port-2 of the PI-CIR. Particularly, the delay line is built in free space and consists of two fiber collimators, a rotatable half-wave plate, a right-angle reflector (M1), a 60-mm linear stage (THORLABS ELL20/M) and a hollow roof prism reflector (M2) installed on it. Hence, the total length in the counter-clockwise direction includes ~6.93 m in fibers and ~0.21 m in air that has a close net-dispersion of about -0.149 ps$^2$. The clockwise and counter-clockwise signals are respectively delivered from 10% extracting ports of OC1 and OC2.

The optical spectra of output signals are measured by an optical spectrum analyzer (OSA, YOKOKAWA AQ6370D). An electrical spectrum analyzer (ESA, AGILENT E4440A) is used to obtain radio frequency (RF) spectra. The repetition rates are tracked by two frequency counters (HEWLETT PACKARD 53181A) that combined with two low-noise photoreceivers (NEW FOCUS 1611). In addition, an intensity noise analyzer (INA, THORLABS PNA1) and a power meter with a fiber-coupled sensor (THORLABS, PM400 and S155C) are employed for testing laser robustness. The environmental changes are monitored by a temperature and humidity logger (THORLABS, TSP01).

To secure optimal coupling efficiency, a beam profiler (THORLABS BP209-IR2/M) is applied to accurately align the scannable delay line with the fiber collimator that couples output light back to the main oscillator. The alignment performance of the delay line is shown in Fig. 2 (a), which reflects the coordinates of angular divergence versus the stage's displacement when the output light reaches the incident plane of the fiber collimator. The angular divergence on the horizontal (X) and the vertical (Y) axis are respectively depicted in Fig. 2 (b) and 2 (c), where the maximum absolute deviations are 0.20 mrad and 0.27 mrad. Their root-mean-square (RMS) values are merely 0.07 mrad and 0.12 mrad. Furthermore, figure 2 (d) illustrates all the deviations projected onto the incident plane of the fiber collimator to indicate the excellent coupling efficiency given that the collimator's numerical aperture (NA) is as large as 0.16 (~160 mrad). This would greatly benefit the high accuracy of the delay line and contribute to the high stability of the laser during repetition rate scanning. Here, the insertion loss of the delay line versus the scanning displacement is measured as shown in Fig. 2 (e), which maintains RMS values of 1.985 dB and 1.984 dB in forward and backward scans, respectively. It is not until the stage approaches its maximum travel of 60 mm that the insertion loss suddenly increases to around 2.8 dB. This might be due to the slight deflection of the stage that introduces a lateral movement of the reflector in synchronization with the stage displacement. Some light could incident on the reflector's edge and be blocked by the aperture as the stage approaches its maximum travel.

3. Experiment results

3.1 Rapid-scanned repetition rate

The bidirectional fiber laser is self-started as the pump power increasing upon 45.5 mW. Two mode-locked pulse trains with orthogonal polarizations are obtained in opposite propagating directions: the scan comb oscillates counter-clockwise and travels along the fast axis of panda fibers; the local comb oscillates clockwise and travels along the slow axis. In particular, the scanning of the fundamental repetition rate in counter-clockwise direction is realized through driving the linear stage of the delay line. The 60-mm maximum travel of the stage provides a maximum optical path difference of 120 mm, which enables a maximum time delay of about 400 ps. This results in a difference frequency $\Delta f_{rep}$ that is approximately linear-positive related to the cavity repetition rate $f_{rep}$. The higher the fundamental repetition rate, the greater the difference frequency. Figure 3 (a) presents the optical spectra during the frequency scan, where the bandwidth of the scan comb (magenta to turquoise lines) is narrower than that of the local comb (aqua green line). Figure 3 (b) summarizes the changes in the central wavelength and 3-dB bandwidth, which fluctuate slightly when the stage moves. The RMS central wavelength of scan comb is 1563.09 nm with a range of 0.17 nm, denoting good coherence with the local comb that centered at 1563.22 nm; while the RMS bandwidth is 0.87 nm with a range of 0.23 nm. Furthermore, the RF spectra during the frequency scan are measured as illustrated in Fig. 3 (c). The signal to noise ratio (SNR) of each component is nearly 80 dB, indicating great stability. Here, the frequency scanning efficiency is around 5.48 kHz/mm within a wide range of ~329 kHz, from ~28.41 MHz to ~28.74 MHz. In contrast to the local comb oscillating at ~29.04 MHz, the difference frequency can be tuned over a wide coverage range up to 633 kHz in the microwave domain. Benefiting from the sharp driving response of resonant piezoelectric motors, the linear stage has a top velocity of 90 mm/s, which enables an unparallelly rapid scanning rate of 493.5 kHz/s.

3.2 Robustness of synchronization between scan and local combs

One of the greatest advantages of a bidirectional mode-locked fiber laser is the good inter-comb synchronization. The impacts attributed to ambient perturbations such as temperature fluctuations are actually equitably imposed on the scan and the local combs that share the same main oscillator. Therefore, the repetition rate of the scan comb is always referenced to that of the local comb. This reference-free capability is promising to release a frequency-scanned laser from complicated active stabilization systems and feedback loops, thereby promoting practical uses beyond the laboratory. Here, the short-term robustness of synchronization between the scan and the local combs is summarized in Fig. 4. Starting from the laser mode-locking initialization and lasting for 1 hour, the tracking record of the two combs' repetition rates is shown in Fig. 4 (a). Their repetition rates first decrease rapidly, next increase sharply, and then slowly climb as the ambient temperature declines — both are always in sync with a difference frequency of ~334686.3 Hz. The standard deviations (SD) of the scan and the local combs are ~3.76 Hz and 3.82 Hz, respectively. Figure 4 (b) depicts the difference frequency, which exhibits excellent robustness with a SD of only 0.10 Hz and a range as small as 0.52 Hz. The corresponding temperature and humidity changes are shown in Fig. 4 (c). The temperature drops by 0.47 °C and the humidity increases by 1% in 1 hour. In this circumstance, the maximal fluctuation of the reference-free difference frequency is only 2% of the 22.43-Hz variation of either the scan or the local comb.

Furthermore, the long-term robustness of synchronization is illustrated in Fig. 5, in which the repetition rates of both combs are continuously tracked for up to 24 hours. Figure 5 (a) presents the overall changes in repetition rates: they are synchronized at all times, experiencing a short descent and climb at the beginning, and then a slow rise and fall in response to temperature drops and climbs. The standard deviations of the scan and the local comb are 81.40 Hz and 82.83 Hz, within a range of 275.19 Hz. The changes in the difference frequency are also recorded in Fig. 5 (b), which gives an RMS value of ~334702 Hz with a standard deviation of 1.54 Hz and a small range of 5.46 Hz. The maximal changes in temperature and humidity are 1.43 °C and 6% as shown in Fig. 5 (c). The 24-hour fluctuation of the difference frequency is likewise reduced to less than 2% of a single comb's fluctuation.

However, two mode hops are found during the 24-hour continuous measurements recorded in Fig. 5(a). This might be caused by unrecorded sudden environmental perturbations other than temperature and humidity, such as unexpected vibrations or pump power fluctuations. The corresponding optical spectra sampled at each 10 seconds are shown in Fig. 6, from which two spectral discontinuities can be found. The first mode hop occurs at approximately the eighth hour. Sudden spectral flutters of the scan comb are observed in the inset of a zoomed-in 3-minute clip reflecting mode hops. The optical spectrum of the local comb also experienced a spectral degradation. Then the hopping spectra return to their original states, and the repetition rates remain synchronized except for a global mutation. This reveals that the proposed laser has a certain anti-interference and self-reconstruction capability. Similarly, the second mode hop happens after the twentieth hour, where the spectrum of the scan comb switched between pulsed and continuous wave states. At this moment, the optical spectrum of the local comb is also compressed into a continuous wave emission. The spectra recovered themselves after this transit operation, however the sampling of the repetition rates is interrupted. Particularly, the second mode hop lasts longer than the first one, showing more severe spectral fluctuations and stronger continuous wave emission. This indicates that the laser has degraded from a stable mode-locked state to a Q-switched or chaotic state for too long. As a result, the frequency counters continuously read repetition rates outside the metering range (0.1 Hz ~ 225 MHz) for a long period of time. Therefore, the control computer continuously receives errors from the frequency counters, thus determining that the laser has entered a chaotic state and the frequency measurement is terminated. Considering that the spectra of the laser are finally rebuilt to stable states, the scan and the local combs are demonstrated to maintain solid synchronization of the mode-locked repletion rates by self-reconstruction throughout the 1-day experiment.

3.3 Relative intensity noise

Apart from the performance of lasing robustness in the frequency domain, the relative intensity noise (RIN) of the bidirectional polarization-multiplexed fiber laser is analyzed. As illustrated in Fig. 7, the scan and the local combs perform equally good on low-frequency flicker noise and high-frequency shot noise. While the scan comb shows higher intensity noise on mid-frequency band from relaxation oscillations, with the worst RIN of around -95 dBc/Hz at 20 kHz. The maximal RMS RIN of the scan and the local comb is around 0.19% and 0.08%, respectively. Either of the two combs experiences most of the incremental noise in the 10 kHz to 30 kHz frequency band. Considering that the gain fiber is bidirectionally pumped by the same pump light source, the associated noise is shared by the two combs. Therefore, the additional intensity noise of the scan comb could be derived from the loss or mechanical vibrations in the scanning branch.

4. Summary

In conclusion, a bidirectional polarization-multiplexed fiber laser is demonstrated to deliver reference-free dual combs with a rapid-scanned rate up to 493.5 kHz/s. The laser is composed of a shared main oscillator and two independent branches, which simultaneously enable a local comb traveling clockwise along with the slow axis of the fibers and a scan comb propagating counter-clockwise by the fast axis. An accurately aligned delay line driven by resonant piezoelectric motors is applied in the scan-comb branch, with angle divergences of 0.07 mrad horizontally and 0.12 mrad vertically, to provide approximately 400-ps time delay. Therefore, as wide as 329-kHz scanning range of repetition rate is realized, indicating a difference frequency reference to the local pulse train up to 633 kHz. Particularly, most of the environmental perturbations are evenly shared by both the two combs. Thus, the laser shows a good robustness either in short-term or long-term operation. The repetition rates of the two combs remain synchronized regardless of changes in ambient temperature and humidity, resulting in maximum variations of 0.52 Hz and 5.46 Hz in 1-hour and 1-day difference frequency measurements, respectively. Despite mode hops, this laser is also demonstrated with the capability to rebuild its original steady state once out of mode-locking. This work explores a potential solution for rapid-scanned and reference-free repetition rate generation that could be practical in frequency heterodyne measurement and laser synchronization.

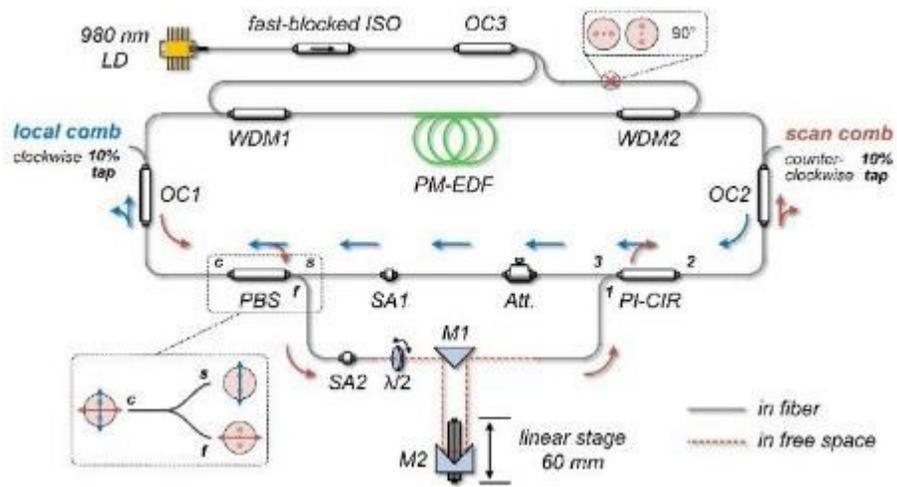

Fig. 1. Schematic diagram of the bidirectional polarization-multiplexed fiber laser: LD, laser diode; ISO, isolator; OC, optical coupler; WDM, wavelength-division multiplexer; PM-EDF, polarization-maintaining erbium-doped fiber; PI-CIR, polarization-independent circulator; SA, saturable absorber; Att., attenuator; M, mirror; PBS, polarizing beam splitter.

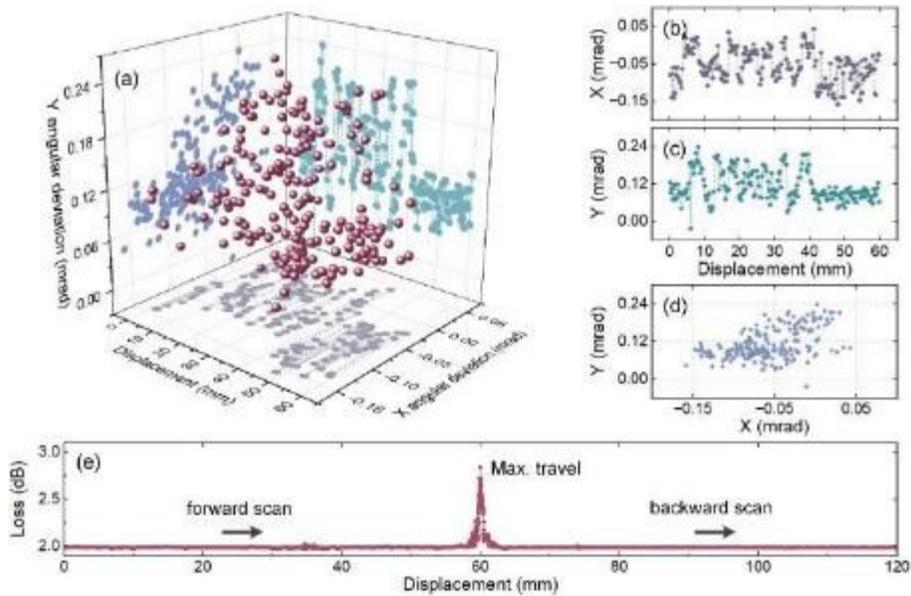

Fig. 2. Performance of the delay line driven by linear piezo stage: (a) overall spread of the output light; (b) horizontal and (c) vertical spread versus displacement; (d) spread on the incidence plane of the collimator; (e) insertion loss introduced by the delay line during scan.

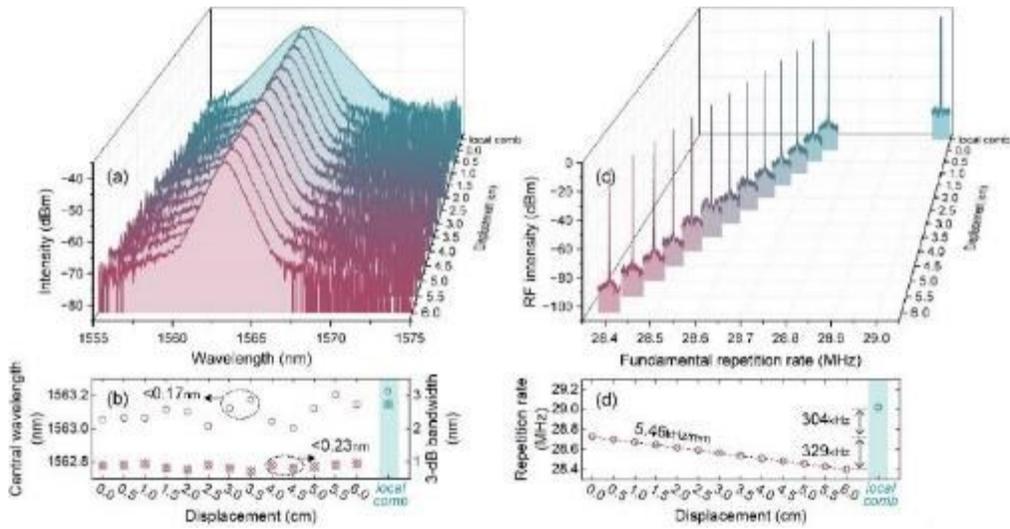

Fig. 3. Scan of repetition rate: (a) optical spectra; (b) changes in central wavelength and 3-dB bandwidth; (c) RF spectra; (d) fundamental repetition rates.

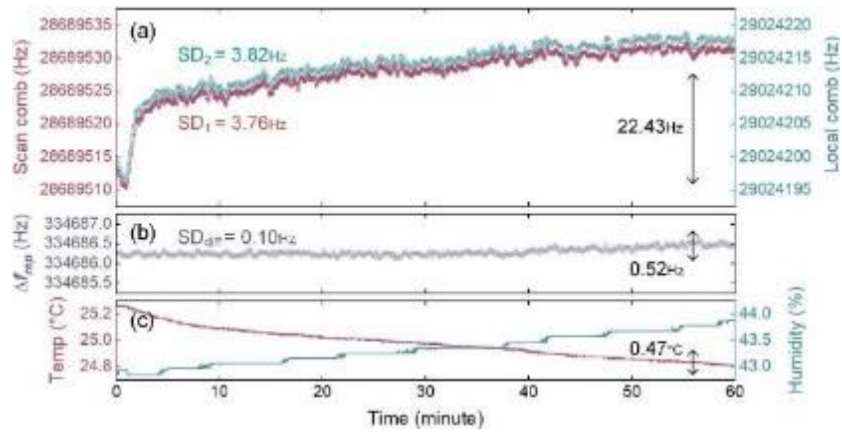

Fig. 4. One-hour robustness of free-running dual combs: (a) repetition rates of clockwise local comb and counter-clockwise scan comb; (b) difference frequency between local and scan combs; (c) fluctuations in ambient temperature and humidity.

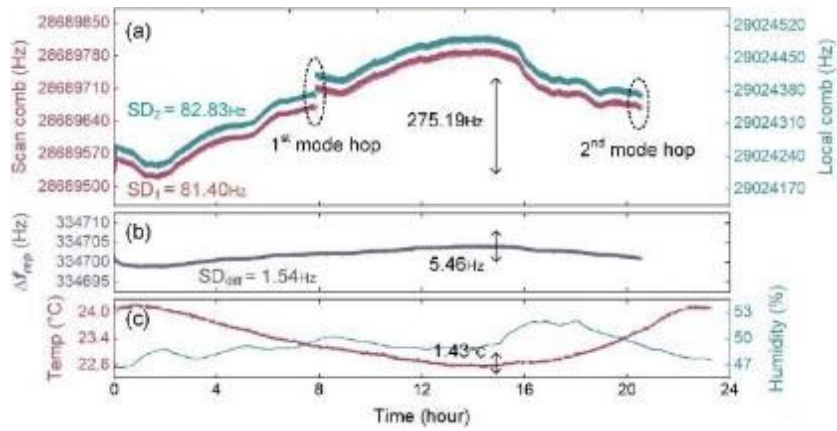

Fig. 5. One-day robustness of free-running dual combs: (a) repetition rates of clockwise local comb and counter-clockwise scan comb; (b) difference frequency between local and scan combs; (c) fluctuations in ambient temperature and humidity.

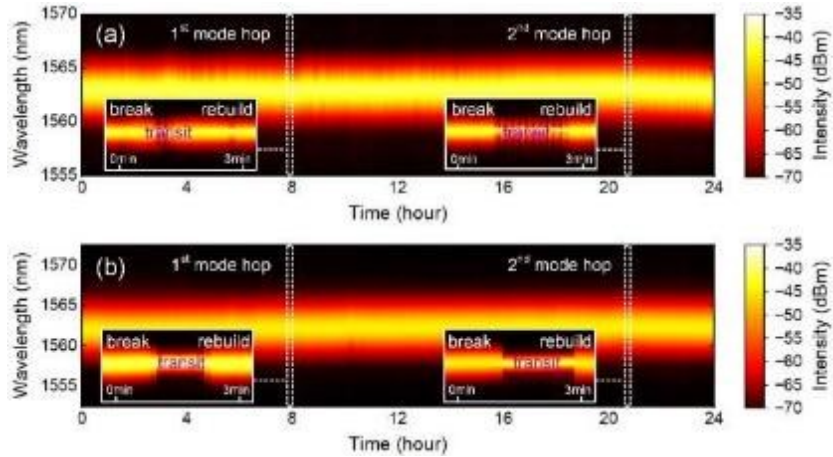

Fig. 6. Mode hops and steady-state reconstruction during 24-hour measurements: (a) the scan comb and (b) the local comb experience two mode hops and then rebuilt to their original states. (inset: enlarged 3-minute spectral clips containing mode hops)

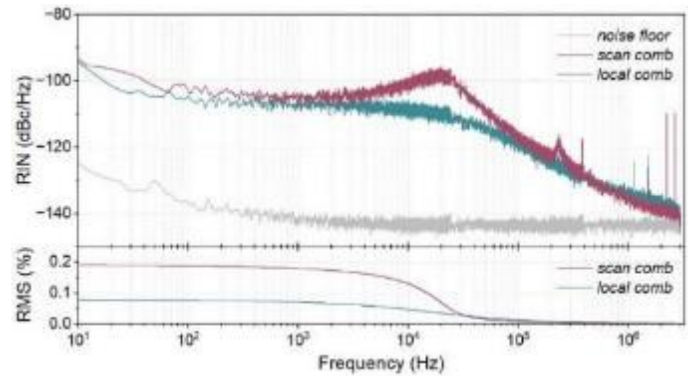

Fig. 7. Relative intensity noise of the photoreceiver's background, the local and the scan comb.